# Measurement of frequency sweep nonlinearity using atomic absorption spectroscopy

*Ningfang Song, Xiangxiang Lu,[①] Xiaobin Xu,\* Xiong Pan, Wei Li, Di Hu, and Jixun Liu*
*School of Instrumentation Science and Opto-electronics Engineering, Beihang University, Beijing 100191, China*
*\*xuxiaobin@buaa.edu.cn*
[①]*luxiangxiang@buaa.edu.cn*

A low cost scheme to determine the frequency sweep nonlinearity using atomic saturated absorption spectroscopy is demonstrated. The frequency modulation rate is determined by directly measuring the interference fringe number and frequency gap between two atomic transition peaks of rubidium atom. Experimental results show that the frequency sweep nonlinearity is ~7.68%, with the average frequency modulation rate of ~28.95 GHz/s, which is in good agreement with theoretical expectation. With this method, the absolute optical frequency and optical path difference between two laser beams are simultaneously measured. This novel technique can be used for applications such as optical frequency sweep nonlinearity correction and real-time frequency monitor.

## 1. Introduction

Diode lasers are widely used in scientific fields ranging from wireless communication [1,2], inverse synthetic aperture lidar detection [3,4] to atomic experiment [5-7]. However, during the frequency sweep process, undesirable nonlinearity occurs and results in significant measurement errors [8,9]. Therefore, precise measurement and feedback control of the frequency sweep nonlinearity is important for high precision applications.

Self-heterodyne interferometry is a conventional method to measure the frequency sweep nonlinearity [10-12]. However, radio frequency signals along with optical modulator and long fiber delay line are needed. On the other hand, wavelength meters, Fabry-Perot etalon (FPE) or frequency comb can be used to monitor the laser frequency and measure the sweep rate of the diode laser [13-16], but the tuning of the laser must be much larger than the free spectral range of the FPE and only gives frequency information at discrete intervals. Another method uses an environmentally isolated reference interferometer to actively correct the frequency sweep nonlinearity [17-19]. The nonlinearity can also be compensated by externally triggering time domain sampling or using post processed resampling algorithms [20,21]. All these methods need expensive and complicated frequency reference.

For high precision applications, tunable diode lasers are usually frequency locked using atomic saturated absorption spectroscopy (SAS) which has excellent frequency stability and accuracy. In this paper we propose and demonstrate a novel and low cost scheme for the measurement of frequency sweep nonlinearity based on the SAS technique. This scheme has several advantages. Firstly, this measurement can be accomplished with conventional laboratory equipment, and do not rely on expensive and bulky wavelength measurement devices to monitor the frequency, thus greatly reducing system complexity and cost. Moreover, this method enables simultaneous determination of frequency sweep rate, absolute optical frequency and optical path difference (OPD) between two beams for one single measurement. This simple technique has potential application in fields such as frequency sweep nonlinearity correction and real-time optical frequency monitor.

## 2. Measurement Principle

For simplicity, the period of the triangular-wave signal used to modulate the laser frequency is defined as $2T_m$. Each period is divided into two parts, the rising period and the falling period, as shown in Fig. 1. The solid curve in upper trace represents the frequency of the signal wave, the dashed curve stands for the frequency of the reference wave, the middle trace is atomic SAS signal of rubidium atom and the solid curve in lower trace corresponds to the periodical beat signals. When the reference wave and the signal wave interfere, the beat signal can be written as [22]

$$I(t) = I_0[1+\eta \cos(2\pi \times \alpha\tau t + 2\pi v_0 \tau - \pi\alpha\tau^2)], \qquad (1)$$

where $I_0$ is the average optical intensity of the beat signal, $\eta$ the fringe contrast, $t$ the sweep time, $\alpha$ the frequency sweep rate, $\tau$ the group time delay given by $\tau = L/c$ where $L$ is the OPD and $c$ is the speed of light, and $v_0$ the average optical frequency. The first term in bracket of Eq. (1) represents the frequency modulation related with the modulation rate and OPD, while the last two terms contribute to the systematic measurement errors.

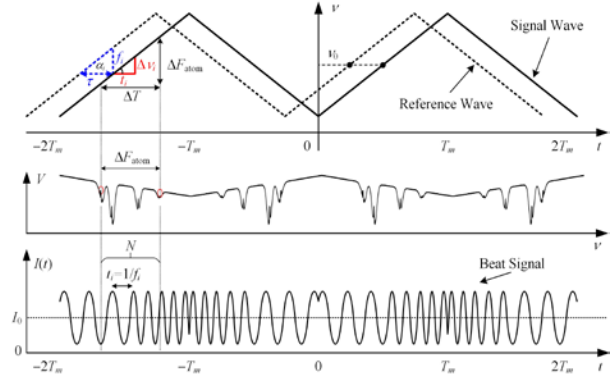

Fig. 1 (color online). Measurement principle of the laser system. Upper trace: scanning waves; middle trace: atomic SAS signal of rubidium atom, $V$ is voltage and $v$ optical frequency; lower trace: beat signals.

From Eq. (1), the beat frequency of the $i$th interference fringe can be written as

$$f_i = \alpha_i \times \tau \quad (i=1,2,...N), \qquad (2)$$

where $N$ is the fringe number, $\Delta T$ ($\Delta T < T_m$) the sweep time corresponding to the two transition peaks in Fig. 1, and $\alpha_i$ the frequency sweep rate during the time interval $t_i$. If $N$ is large enough, it is reasonable to assume that the sweep rate is constant for each time interval $t_i$ with the relation $\sum_{i=1}^{i=N} t_i = \Delta T$. On the other hand, for each time interval $t_i$, the optical frequency increases $\Delta v_i = \alpha_i \times t_i$. By adding each minor frequency increment during $\Delta T$ yields $\sum_{i=1}^{i=N} \Delta v_i = \Delta F_{atom}$, where $\Delta F_{atom}$ is the fixed frequency gap of the two transition peaks of alkali atoms. Note that for certain atoms $\Delta F_{atom}$ is a constant value and can be precisely known. This typical value is on the order of GHz and immune to environmental perturbations.

The frequency modulation rate and the absolute optical frequency can be calculated using Eq. (3). The fringe number $N$ of the beat signal is counted with peak finding algorithms and each beat frequency can be determined using the peak location related scanned time.

$$\begin{aligned}\alpha_i &= f_i / \tau = f_i \times \Delta F_{atom} / N \\ v_i &= v_0 + i \times \Delta F_{atom} / N \quad (i=1,2,...N).\end{aligned} \qquad (3)$$

Furthermore, the delay time and OPD can also be determined as follows

$$\tau = N / \Delta F_{atom}, \quad L = cN / \Delta F_{atom}. \qquad (4)$$

The spatial OPD resolution is $\Delta L = c / \Delta F_{atom}$, consistent with that reported in [21].

## 3. Experimental Results

According to the measurement principle described in section 2, schematic of the experimental apparatus is depicted in Fig. 2 to measure the frequency sweep nonlinearity. The tunable laser is an external cavity diode laser (ECDL, Toptica DL Pro) operating at 780.24 nm, which corresponds to resonant transitions of rubidium (Rb) atom. In principle any two of the transition peaks can be used as the frequency reference, and the transition $5S_{1/2}F=2 \rightarrow 5P_{3/2}F'=2,3$ crossover of $^{87}Rb$ to the transition $5S_{1/2}F=1 \rightarrow 5P_{3/2}F'=1,2$ of $^{87}Rb$ (the frequency range $\Delta F_{atom}$ is 6.623 GHz) is used in this experiment, where F and F' denote the ground states and excited states respectively. The laser frequency is modulated with the triangular wave by driving the piezoelectric transducer (PZT) with a repetition rate of 1 Hz and a peak-to-peak scanning voltage of 30 V, corresponding to overall scanned frequency range of 14 GHz. The points A, B and C represent the measurement locations of the frequency modulation rate and absolute optical frequency, the SAS and the beat signal respectively.

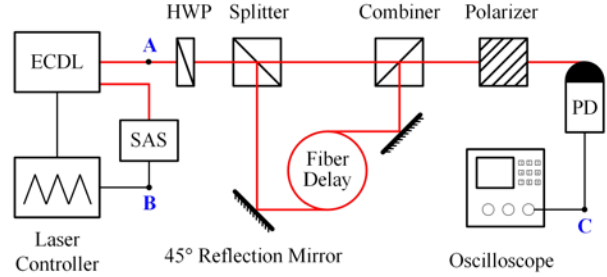

Fig. 2 (color online). Schematic of experimental apparatus. Red line: optical signal; black line: electric signal.

The laser frequency is locked onto one of the transition lines of $^{87}Rb$ atom with SAS technique. The ECDL output light is split into two beams by utilizing a half wave plate (HWP) and splitter. A polarization maintaining (PM) fiber is used as the optical time delay. The two beams are recombined by the combiner and linearly polarized by a Glan-Taylor prism. The beat signal is received by a high-speed photodetector (PD, Newport 1554B), and the converted electronic signal (70 mV for each beam) is sent to an oscilloscope (Tektronix MDO3104) for real-time signal acquisition and analysis. The sampling rate is set to be 1 MS/s (million samples per second) with a total data acquisition length of 1 MS and a sampling time of 10 s.

The experimental results are shown in Fig. 3, with atomic SAS signal in Fig. 3 (a) and beat signal in Fig. 3 (b). The time interval $\Delta T$ is determined to be ~0.226 s, corresponding to the two peaks of atomic signal in Fig. 3 (a) (red elliptical circle with dotted-line). The interference fringe number $N$ is 178 by using peak finding algorithms. Clearly periodic fringes can be observed from the inset (red box) in Fig. 3 (b).

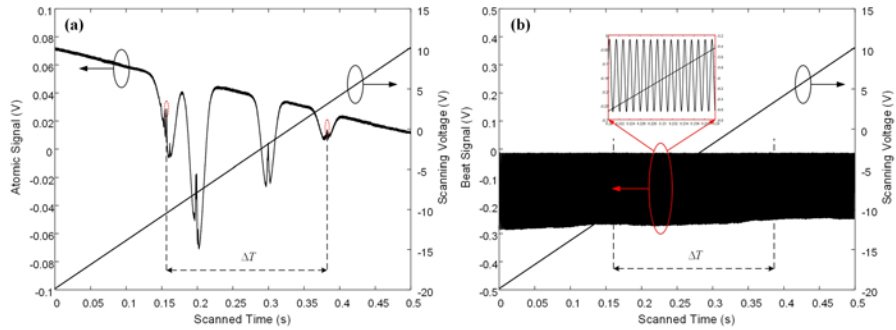

Fig. 3 (color online). Experimental results. (a) Atomic signal at point B in Fig. 2. (b) Beat signal at point C in Fig. 2.

The frequency modulation rate and absolute optical frequency (measured at point A in Fig. 2) are calculated using Eq. (3), which is shown in Fig. 4(a). The frequency modulation rate is about 28.95±2.2 GHz/s, implying a frequency sweep nonlinearity of ~7.68%. As expected, the frequency modulation rate is not constant but shows a weak linear relationship with the scanned time. Furthermore, clear oscillations of the frequency modulation rate over time is visible, consistent with that reported in [23]. To further confirm the validation of this measurement, we use a commercial wavelength meter (HighFinesse WS7) to record the frequency variation under same experimental conditions, which is shown in Fig. 4(b). The true frequency modulation rate is about 34.23±21.6 GHz/s, which means the measurement based on the SAS technique is valid. This can also be verified by comparing the two figures in Fig. 4, both the modulation rate and optical frequency do have similar trends. However, oscillations of the true frequency modulation rate are also observed, and is one order larger than that of with the SAS technique. The reason for this large oscillations is unknown yet and subsequent study is under way.

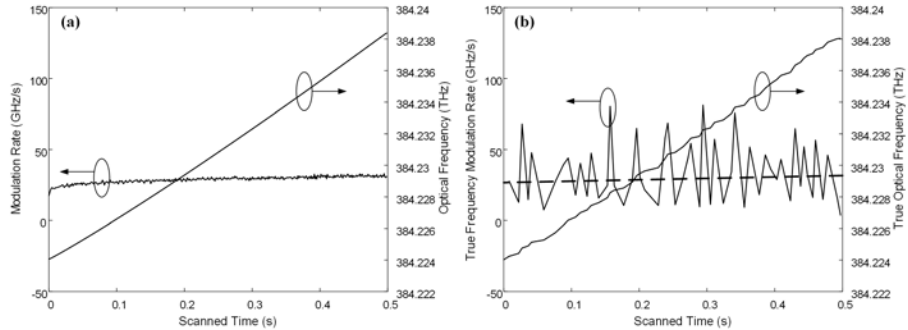

Fig. 4. The frequency modulation rate and absolute optical frequency over scanned time. (a) Measured with the SAS technique. (b) Measured with the commercial wavelength meter (HighFinesse WS7).

Moreover, this technique can also be used to simultaneously measure the absolute optical frequency and OPD for one single run. The absolute optical frequency changes from 384.228 THz to 384.2346 THz with a frequency resolution of 37.2 MHz in the 0.236-s scanned time. It seems linear frequency tuning but essentially contains small frequency fluctuation, which can be seen in Fig. 4(b). This high frequency resolution is comparable to that of conventional wavelength meters, but features with low cost and system simplicity, which indicates potentially portable application for field frequency monitor.

Using Eq. (4) the OPD (delay time) is calculated to be 8.057 (26.876 ns), which is in good agreement with the true OPD (delay time) of 8.12 m (27.085 ns). This measurement error of 0.77% is mainly due to the fringe number uncertainty $\Delta N$, which essentially resulted from the ambiguity of precise scanned time, which corresponds to atomic transition peak locations. For experimental conditions, the spatial OPD (delay time) resolution $\Delta L$ ($\Delta \tau$) is calculated to be 4.5 cm (0.15ns).

### 4. Discussion

The frequency modulation rate nonlinearity arises from several reasons. Firstly, the measurement is affected by vibration and airflow fluctuation. According to [23], drifts and vibrations occurring along the optical path will be magnified by a factor of $\Omega = v_0 / \Delta v$ (~$5.8 \times 10^4$). Small vibrations and drift errors can result in significant effects. On the other hand, because the optical path length of dispersive elements changes during the frequency sweep, elements such as the splitter, combiner and PM fiber induce considerable nonlinearity for the ~ 8 m delay length. Moreover, direct modulation of the laser cavity with a triangular wave also induces frequency sweep nonlinearity, which is due to the fact that the laser frequency is inversely proportional to the cavity length. Therefore, even a linear sweep of the laser cavity still result in nonlinearity of the frequency sweep.

To get a more straightforward knowledge on the frequency modulation rate nonlinearity, a 2nd order polynomial fitting of the modulation rate over scanned time is performed, as illustrated in Fig. 5. The fitting can be expressed as $\alpha(t) = \alpha_2 t^2 + \alpha_1 t + \alpha_0$, where $\alpha(t)$ is the measured chirp rate. After some calculations, the corresponding frequency tuning coefficients are determined to be $\alpha_2$ = −0.604±0.0.0948 GHz/s$^3$, $\alpha_1$ = 1.937±0.0855 GHz/s$^2$, and $\alpha_0$ = 29.56±0.125 GHz/s respectively. The former two terms represent the 3rd order and 2nd order frequency modulation of the frequency sweep process. The fitting results are in consistent with that of [24].

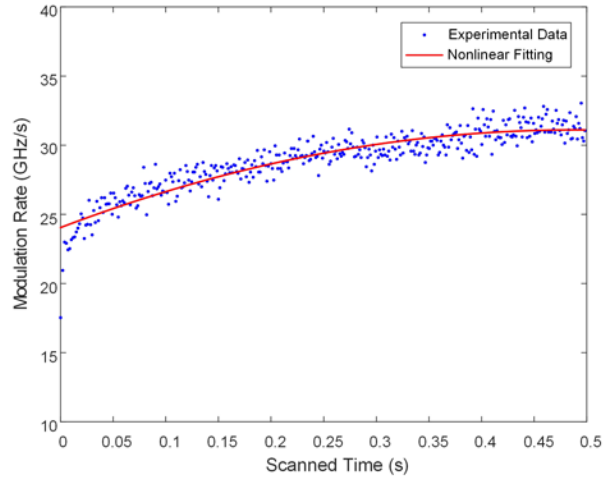

Fig. 5 (color online). Nonlinear fitting of the frequency modulation rate over scanned time.

The technique using atomic saturated absorption spectroscopy enables simultaneous measurement of the frequency modulation rate and its nonlinearity, OPD and absolute optical frequency. All these measurements are based on routine laboratory equipment and no bulky or expensive wavelength meters are needed, greatly reducing system complexity and cost.

### 5. Conclusion

In this paper, a novel and low cost scheme to measure the frequency sweep nonlinearity is proposed and demonstrated. The principle of using atomic saturated absorption spectroscopy to determine the frequency modulation rate and the sweep nonlinearity is presented. By determining the interference fringe number in scanned time, and the frequency gap of atomic transition peaks of rubidium atom, the frequency modulation rate is measured to be 28.95±2.2 GHz/s with a nonlinearity of ~7.68%, which is in good agreement with theoretical expectation. Moreover, this technique enables simultaneous measurement of the OPD with a measurement error of 0.77% and a spatial resolution of 4.5 cm. The absolute optical frequency is monitored with a resolution of 37.2 MHz. This technique can be used for frequency sweep nonlinearity correction and real-time frequency monitor.

## Acknowledgements

The authors would like to thank Fuyu Gao and Haoshi Zhang for helping to measure the fiber coil length and Wei Cai for useful discussions.